# Chapter 1
# Partitions of a Finite Partially Ordered Set

Pietro Codara
Dipartimento di Informatica e Comunicazione
Università degli Studi di Milano, Italy

**Abstract** In this paper, we investigate the notion of partition of a finite partially ordered set (poset, for short). We will define three different notions of partition of a poset, namely, monotone, regular, and open partition. For each of these notions we will find three equivalent definitions, that will be shown to be equivalent. We start by defining partitions of a poset in terms of fibres of some surjection having the poset as domain. We then obtain combinatorial characterisations of such notions in terms of blocks, without reference to surjection. Finally, we give a further, equivalent definition of each kind of partition by means of analogues of equivalence relations.

## 1.1 Introduction

A partition of a set $A$ is a collection of nonempty, pairwise disjoint subsets, often called *blocks*, whose union is $A$. Equivalently, partitions can be defined by means of *equivalence relations*, by saying that a partition of a set $A$ is the set of equivalence classes of an equivalence relation on $A$. A third definition of a partition can be given in terms of *fibres* of a surjection: a partition of a set $A$ is the set $\{f^{-1}(y) \mid y \in B\}$ of fibres of a surjection $f : A \to B$.

In this paper, we investigate the notion of partition of a finite poset. Our goal is to find the analogues of the three definitions of partition of a set mentioned above, in terms of blocks, relations, and fibres, respectively. For the rest of this paper, we shall omit the term finite, as we only deal with finite posets; 'poset partition' means 'partition of a poset'.

We begin our study of poset partitions with the notion of partition given in terms of fibres. Some background in category theory, and a few preliminary results on two different categories having posets as objects (Section 1.2), will allow us to identify three kind of morphisms appropriate to introduce our first definition. In Section 1.3, we introduce monotone, regular, and open partitions, according to the morphisms we are considering.

The definitions given in Section 1.3 need to mention morphism to describe what a poset partition is. To investigate the combinatorial structure of a poset partition, it is useful to have intrinsic notions of poset partitions making no reference to morphisms. In Section 1.4 we obtain such a combinatorial characterisation of each kind of partition.

In Section 1.5, we investigate the analogues of the definition of partition of a set given by means of equivalence relations. Again, we obtain characterisations of monotone, regular, and open partitions in this terms.

This piece of work contains a revisited and extended version of some of the results obtained in [3], where the author defines and analyses the notions of monotone and regular partitions of a poset, and in [4], where open partitions are introduced.

**Acknowledgment.** The author wishes to thank Prof. O. D'Antona, and Dr. V. Marra for their helpful comments and suggestions, and for the many discussions on this topic.

## 1.2 Background, and Preliminary Results

When one defines a partition of a set in terms of fibres, one makes use of a special class of morphisms of the category[1] Set of sets and functions. In fact, such definition exploits the notion of surjection, which can be shown to coincide in Set with the notion of *epimorphism*. Recall that an epimorphism in a category is a morphism $f : A \to B$ that is right-cancellable with respect to composition: whenever $h \circ f = k \circ f$, for $h, k : B \to C$, we have $h = k$. The category-theoretic dual of the notion of epimorphism is *monomorphism*. In Set, monomorphisms coincide with injections. The well-known fact that each function between sets factorises (in an essentially unique way) as a surjection followed by an injection can be reformulated in categorical terms by saying that the classes of epimorphisms and monomorphism form a factorisation system for Set, or, equivalently, that (epi,mono) is a factorisation system for Set. Epimorphism and factorisation systems will play a key role in the following.

First, consider the category Pos of posets and *order-preserving maps* (also called *monotone maps*), *i.e.*, functions $f : P \to Q$, with $P$, $Q$ posets such that $x \leqslant y$ in $P$ implies $f(x) \leqslant f(y)$ in $Q$, for each $x, y \in P$. In Pos, (epi,mono) is not a factorisation system; to obtain one we need to isolate a subclass of epimorphisms, called regular epimorphisms. A morphism $e : B \to C$ in an arbitrary category is a *regular epimorphism* if and only if there exists a pair $f, g : A \to B$ of morphisms such that

---

[1] For background on category theory we refer, *e.g.*, to [1].

(1) $e \circ f = e \circ g$,

(2) for any morphism $e' : B \to C'$ with $e' \circ f = e' \circ g$, there exists a unique morphism $\psi : C \to C'$ such that $e' = \psi \circ e$.

Regular epimorphisms are epimorphisms, as one easily checks. While in Set regular epimorphisms and epimorphisms coincide, that is not the case in Pos. The dual notion of regular epimorphism (obtained by reversing arrows in the above statement) is *regular monomorphism*. It can be shown (see, *e.g.*, [3, Proposition 2.5]) that (regular epi,mono) is a factorisation system for the category Pos. A second factorisation system for Pos is given by the classes of epimorphisms and regular monomorphisms. In other words, each order-preserving map between posets factorises in an essentially unique way both as a regular epimorphism followed by a monomorphism, and as an epimorphism followed by a regular monomorphism.

The existence of two distinct factorisation systems in Pos leads us to introduce two different notions of partition of a poset, one based on the use of epimorphisms, the other based on the use of regular epimorphisms.

Our next step is to characterise regular epimorphisms.

**Notation.** If $\pi$ is a partition of a set $A$, and $a \in A$, we denote by $[a]_\pi$ the block of $a$ in $\pi$. When no confusion is possible, we shall write $[a]$ instead of $[a]_\pi$. Further, let us stress our usage of different symbols for representing different types of binary relations. The symbol $\leqslant$ denotes the partial order relation between elements of a poset. A second symbol, $\lhd$, represents the associated covering relation. Finally, the symbol $\lesssim$ denotes *quasiorder relations*, sometimes called *preorders*, *i.e*, reflexive and transitive relations.

**Definition 1.1 (Blockwise quasiorder).** Let $(P, \leqslant)$ be a poset and let $\pi$ be a partition of the set $P$. For $x, y \in P$, $x$ *is blockwise under $y$ with respect to $\pi$*, written $x \lesssim_\pi y$, if and only if there exists a sequence $x = x_0, y_0, x_1, y_1, \ldots, x_n, y_n = y \in P$ satisfying the following conditions.

(1) For all $i \in \{0, \ldots, n\}$, $[x_i] = [y_i]$.

(2) For all $i \in \{0, \ldots, n-1\}$, $y_i \leqslant x_{i+1}$.

Observe that the relation $\lesssim_\pi$ in Definition 1.1 indeed is a quasiorder. In fact, if $x \leqslant y$ and $y \leqslant z$ for $x, y, z \in P$, then there exist two sequences $x = x_0, y_0, x_1, y_1, \ldots, x_n, y_n = y$ and $y = y_{n+1}, z_{n+1}, y_{n+2}, z_{n+2}, \ldots, y_{n+m}, z_{n+m} = z$ satisfying (1) and (2), and a sequence $x = x_0, y_0, x_1, y_1, \ldots, x_n, y_n = y_{n+1}, z_{n+1}, y_{n+2}, z_{n+2}, \ldots, y_{n+m}, z_{n+m} = z$ satisfying (1) and (2), too. Thus, $x \lesssim_\pi z$ and the relation $\lesssim_\pi$ is transitive. The reflexivity of $\lesssim_\pi$ results trivially.

The definition of blockwise quasiorder allows us to isolate a special kind of order-preserving map.

**Definition 1.2 (Fibre-coherent map).** Consider two posets $P$ and $Q$. Let $f : P \to Q$ be a function, and let $\pi_f = \{f^{-1}(q) | q \in f(P)\}$ be the set of fibres of $f$. We say $f$ is a *fibre-coherent map* whenever for any $p_1, p_2 \in P$, $f(p_1) \leqslant f(p_2)$ if and only if $p_1 \lesssim_{\pi_f} p_2$.

**Proposition 1.1.** *In* Pos, *regular epimorphisms are precisely fibre-coherent surjections.*

*Proof.* ($\Rightarrow$) Let $(P, \leqslant_P)$ and $(Q, \leqslant)$ be posets, let $e : P \to Q$ be a regular epimorphism and let $\pi_e = \{e^{-1}(q) | q \in Q\}$. Since $e$ is epi, it is an order-preserving surjection. Moreover, by the definition of regular epimorphism, there exists a pair $f, g : R \to P$ of morphisms such that $e \circ f = e \circ g$.

Suppose, by way of contradiction, that $e$ is not fibre-coherent. If $x \lesssim_{\pi_e} y$ for some $x, y \in P$, then there exists a sequence $x = x_0, y_0, x_1, y_1, \ldots, x_n, y_n = y \in P$ satisfying conditions (1) and (2) in Definition 1.1. For such a sequence, since $e$ is order-preserving, we have $e(x) = e(x_0) = e(y_0) \leqslant e(x_1) = e(y_1) \leqslant \cdots \leqslant e(x_n) = e(y_n) = e(y)$. Thus, to satisfy the *absurdum hypothesis* there must exist $p_1, p_2 \in P$, with $e(p_1) = q_1$ and $e(p_2) = q_2$, such that $q_1 \leqslant q_2$ but $p_1 \not\lesssim_{\pi_e} p_2$. Note that $p_1$ and $p_2$ must be incomparable, and that $q_1 \neq q_2$.

Case (i). Suppose $q_1 \lhd q_2$, where $\lhd$ is the covering relation induced by $\leqslant$. Consider the poset $Q'$ having $Q$ as underlying set, endowed with the relation $\leqslant'$ obtained by removing from $\leqslant$ the pair $(q_1, q_2)$. In other words, the only difference between $\leqslant'$ and $\leqslant$ is that $q_1 \leqslant q_2$, but $q_1 \not\leqslant' q_2$. Since $q_1 \lhd q_2$, removing $(q_1, q_2)$ from $\leqslant$ does not impair transitivity and $\leqslant'$ indeed is a partial order.

Now, consider the function $e' : P \to Q'$ that coincides with $e$ on the underlying sets. We want to show that $e'$ is order-preserving. For this, let $x, y \in P$. It suffices to consider two cases only: $e(x) = q_1$ and $e(y) = q_2$, and viceversa. In any other case, $e'$ preserves order just because $e$ does. Suppose, without loss of generality, $x \in e^{-1}(q_1)$ and $y \in e^{-1}(q_2)$. Then, $x \not\leqslant_P y$, for else the chain $p_1, x, y, p_2$ would satisfy conditions (1) and (2) in Definition 1.1, contradicting $p_1 \not\lesssim_{\pi_e} p_2$. Moreover, $y \not\leqslant_P x$, because $e$ is order preserving. Thus, for each $x \in e^{-1}(q_1)$ and $y \in e^{-1}(q_2)$, $x$ and $y$ are incomparable. Summing up, $e'$ is order preserving. Since $e'$ coincides with $e$ on the underlying sets, we obtain $e' \circ f = e' \circ g$.

Since $e$ is a regular epimorphism, by definition, we can find a unique morphism $\psi : Q \to Q'$ such that the diagram in Figure 1.1 commutes. Take $x, y \in P$ such that $e(x) = e'(x) = q_1$ and $e(y) = e'(y) = q_2$. Thus, we should have $\psi(q_1) = q_1$ and $\psi(q_2) = q_2$ but, by hypothesis, $q_1 \leqslant q_2$ and $q_1 \not\leqslant' q_2$, and such a $\psi$ would not be order-preserving. Since $f$ and $g$ satisfy $e \circ f = e \circ g$ but are otherwise arbitrary, this would contradict the fact that $e$ is a regular epimorphism. Therefore, $e$ has to be fibre-coherent.

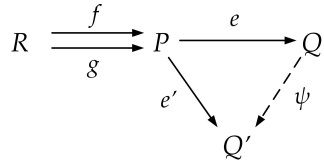

**Fig. 1.1** Proof of Proposition 1.1.

*Case* (ii). Suppose $q_1 \not\vartriangleleft q_2$. Then there exists a sequence $k_1, k_2, \ldots, k_u \in Q$ such that $q_1 = k_1 \vartriangleleft k_2 \vartriangleleft \cdots \vartriangleleft k_u = q_2$. Let $x_1, x_2, \ldots, x_u \in Q$ be such that $x_i \in e^{-1}(k_i)$, for each $i \in \{1, u\}$, and suppose $x_1 \lesssim_{\pi_e} x_2 \lesssim_{\pi_e} \cdots \lesssim_{\pi_e} x_u$. Since $p_1 \in e^{-1}(k_1)$ and $p_2 \in e^{-1}(k_u)$ imply $p_1 \lesssim_{\pi_e} x_1$ and $x_u \lesssim_{\pi_e} p_2$, then, by transitivity, $p_1 \lesssim_{\pi_e} p_2$, contradicting our hypothesis. Thus, there exists an index $j$ such that $k_j \vartriangleleft k_{j+1}$, but $x_j \not\lesssim_{\pi_e} x_{j+1}$. The proof follows now the same steps of *Case* (i), with $x_j$ and $x_{j+1}$ playing the role of $p_1$ and $p_2$, respectively.

($\Leftarrow$) Let $(P, \leqslant_P)$ and $(Q, \leqslant)$ be posets, and let $e : P \to Q$ be a fibre-coherent surjection. Consider the poset $R \subseteq P \times P$, having underlying set $\{(r_1, r_2) \in P \times P \mid e(r_1) = e(r_2)\}$, endowed with the order $\leqslant_R$ defined by $(r_1, r_2) \leqslant_R (s_1, s_2)$ if and only if $r_1 \leqslant_P s_1$ and $r_2 \leqslant_P s_2$. Let $f, g : R \to P$ be the projection functions of $R$, *i.e.*, $f$ and $g$ are the order-preserving maps such that, for each $r = (r_1, r_2) \in R$, $f(r) = r_1$, $g(r) = r_2$. Clearly, $e \circ f = e \circ g$.

We need to show that $e$ is a regular epimorphism. Consider a poset $(Q', \leqslant')$ and an order-preserving map $e' : P \to Q'$ such that $e' \circ f = e' \circ g$. Note that, for each $q \in Q$, if $x, y \in e^{-1}(q)$, there exists $r \in R$ such that $f(r) = x$ and $g(r) = y$. Thus, from $e' \circ f = e' \circ g$ follows that $e'(x) = e'(y)$. Since $e$ is a surjection, we can construct a map $\psi : Q \to Q'$ by setting $\psi(q) = e'(x)$ for some $x \in e^{-1}(q)$, where $q \in Q$.

Let now $q_1, q_2 \in Q$ with $q_1 \leqslant q_2$, and let $x_1, x_2 \in P$ be such that $e(x_1) = q_1$, $e(x_2) = q_2$. By Definition 1.2, we have $x_1 \lesssim_{\pi_e} x_2$. Thus, by Definition 1.1 there exists a sequence $y_0, z_0, y_1, z_1, \ldots, y_n, z_n \in P$ with $x_1 = y_0$ and $x_2 = z_n$ such that $e(y_i) = e(z_i)$, for $i = 0, \ldots, n$, and $z_j \leqslant_P y_{j+1}$, for $j = 0, 1, \ldots, n-1$. Moreover, from $e' \circ f = e' \circ g$ it follows that $e'(y_i) = e'(z_i)$, and, since $e'$ is order-preserving, we have $e'(z_j) \leqslant' e'(y_{j+1})$. Thus, $e'(y_0) = e'(z_0) \leqslant' e'(y_1) = e'(z_1) \leqslant' \cdots \leqslant' e'(y_n) = e'(z_n)$. Therefore, we have $\psi(q_1) = e'(x_1) \leqslant \psi(q_2) = e'(x_2)$ and $\psi$ is order-preserving. The morphism $\psi$ is now well defined and, by construction, satisfies $e'(x) = \psi(e(x))$ for all $x \in P$.

Let $\psi'$ be another map from $Q$ to $Q'$, $\psi \neq \psi'$, and let $\overline{q}$ be an element of $Q$ such that $\psi(\overline{q}) \neq \psi'(\overline{q})$. Since $e$ is surjective, there exists $x \in P$ such that $e(x) = \overline{q}$. Then from $\psi'(\overline{q}) \neq \psi(\overline{q})$ and $e'(x) = \psi(\overline{q})$ we have $\psi'(\overline{q}) \neq e'(x)$ and $\psi' \circ e \neq e'$. Hence, $\psi : Q \to Q'$ is the unique function such that $\psi \circ e = e'$. Summing up, for an arbitrary morphism $e' : P \to Q'$ such that $e' \circ f = e' \circ g$, there exists a unique order preserving map $\psi : Q \to Q'$ such that $\psi \circ e = e'$, *i.e.*, $e$ is a regular epimorphism. □

The second category we are going to consider is the category OPos of posets and open maps. Such maps arise naturally in the investigation of intuitionistic logic; cf. the notion of p-morphisms of Kripke frames, *e.g.* in [2]. An order-preserving function $f : P \to Q$ between posets is called *open* if whenever $f(u) \geqslant v'$ for $u \in P$ and $v' \in Q$, there is $v \in P$ such that $u \geqslant v$ and $f(v) = v'$. One can check that epimorphisms in OPos are surjective open maps, and monomorphisms are injective open maps. As in Set, in OPos each epimorphism is a regular epimorphism. Further, OPos admits an (epi,mono) factorisation system. In the next section, we will see that surjective open maps in OPos induce a third kind of partition of a poset.

## 1.3 Partitions as Sets of Fibres

Poset partitions can be defined in terms of fibres. From the notions of epimorphism and regular epimorphism in Pos, we derive immediately the two following definitions.

**Definition 1.3 (Monotone partition).** A *monotone partition* of a poset $P$ is a poset $(\pi_f, \leqslant)$, where $\pi_f$ is the set of fibres[2] of an order-preserving surjection $f : P \to Q$, for some poset $Q$, and $\leqslant$ is the partial order on $\pi_f$ defined by

$$f^{-1}(q_1) \leqslant f^{-1}(q_2) \text{ if and only if } q_1 \leqslant q_2, \tag{1.1}$$

for each $q_1, q_2 \in Q$.

**Definition 1.4 (Regular partition).** A *regular partition* of a poset $P$ is a poset $(\pi_f, \leqslant)$, where $\pi_f$ is the set of fibres of a fibre-coherent surjection $f : P \to Q$, for some poset $Q$, and $\leqslant$ is the partial order on $\pi_f$ defined by

$$f^{-1}(q_1) \leqslant f^{-1}(q_2) \text{ if and only if } q_1 \leqslant q_2, \tag{1.2}$$

for each $q_1, q_2 \in Q$.

*Remark 1.1.* Since a fibre-coherent map is order-preserving, it follows immediately that each regular partition of a poset is a monotone partition.

Consider now the category OPos.

**Definition 1.5 (Open partition).** An *open partition* of a poset $P$ is a poset $(\pi_f, \leqslant)$, where $\pi_f$ is the set of fibres of a surjective open map $f : P \to Q$, for some poset $Q$, and $\leqslant$ is the partial order on $\pi_f$ defined by

$$f^{-1}(q_1) \leqslant f^{-1}(q_2) \text{ if and only if } q_1 \leqslant q_2, \tag{1.3}$$

for each $q_1, q_2 \in Q$.

*Remark 1.2.* One can check that an open map is fibre-coherent. It follows that each open partition of a poset is a regular partition, whence a monotone one.

There are regular partitions that are not open, and monotone partitions that are not regular; cf. Example 1.1.

## 1.4 Partitions as Partially Ordered Sets of Blocks

For each definition in the previous section, we give a new definition in terms of partially ordered blocks without mentioning morphisms. We prove, for each notion of partition, that the two kinds of definition describe exactly the same objects.

---

[2] Note that, since $f$ is surjective, $\pi_f$ is a partition of the underlying set of $P$.

**Definition 1.6 (Monotone partition).** A *monotone partition* of a poset $P$ is a poset $(\pi, \preccurlyeq)$ where

(i) $\pi$ is a partition of the underlying set of $P$,
(ii) for each $p_1, p_2 \in P$, $p_1 \leq p_2$ implies $[p_1] \preccurlyeq [p_2]$.

**Theorem 1.1.** *Definitions 1.3 and 1.6 are equivalent.*

*Proof.* (Definition 1.6 $\Rightarrow$ Definition 1.3). Let $\pi$ be a partition of the underlying set of a poset $P$, and let $\preccurlyeq$ be a partial order on $\pi$ satisfying (ii). Consider the projection map $f : P \to \pi$ which sends each element of $P$ to its block in $\pi$. Since a partition does not have empty blocks, $f$ is a surjection. By (ii), $f$ is order-preserving. Since, by construction, $f^{-1}(B) = B$ for each $B \in \pi$, the partial order $\preccurlyeq$ satisfies (1.1). Thus, $(\pi, \preccurlyeq)$ is a monotone partition of $P$ according to Definition 1.3.

(Definition 1.3 $\Rightarrow$ Definition 1.6). Let $f : P \to Q$ be an order-preserving surjection, and let $(\pi_f, \preccurlyeq)$ be a monotone partition of $P$, according to Definition 1.3. Since $f$ is surjective, $\pi_f$ is a partition of the underlying set of $P$. Consider $p_1, p_2 \in P$ such that $p_1 \leq p_2$. Since $f$ is order-preserving, $f(p_1) \leq f(p_2)$ holds and, by (1.1), $[p_1] = f^{-1}(f(p_1)) \preccurlyeq [p_2] = f^{-1}(f(p_2))$. We have so proved (ii). Thus, $(\pi_f, \preccurlyeq)$ is a monotone partition of $P$ according to Definition 1.6.

**Corollary 1.1.** *Let $(\pi, \preccurlyeq)$ be a monotone partition of a poset $P$. Then, for each $p_1, p_2 \in P$,*

$$[p_1] = [p_2] \text{ if and only if } (p_1 \lesssim_\pi p_2 \text{ and } p_2 \lesssim_\pi p_1). \tag{1.4}$$

*Proof.* ($\Rightarrow$) Directly from Definition 1.1.

($\Leftarrow$) Let $p_1, p_2 \in P$ be such that $p_1 \lesssim_\pi p_2$ and $p_2 \lesssim_\pi p_1$. Then, there exist two sequences $p_1 = x_0, y_0, x_1, y_1, \ldots, x_n, y_n = p_2$ and $p_2 = z_0, w_0, \ldots, z_m, w_m = p_1$ of elements of $P$ satisfying Conditions (1) and (2) in Definition 1.1, with respect to $\pi$. By Condition (ii) in Definition 1.6, we have $[p_1] = [x_0] = [y_0] \preccurlyeq [x_1] = [y_1] \preccurlyeq \cdots \preccurlyeq [x_n] = [y_n] = [p_2]$, and $[p_2] = [z_0] = [w_0] \preccurlyeq [z_1] = [w_1] \preccurlyeq \cdots \preccurlyeq [z_m] = [w_m] = [p_1]$. Thus, $[p_1] = [p_2]$.

**Definition 1.7 (Regular partition).** A *regular partition* of a poset $P$ is a poset $(\pi, \preccurlyeq)$ where

(i) $\pi$ is a partition of the underlying set of $P$,
(ii) for each $p_1, p_2 \in P$, $p_1 \lesssim_\pi p_2$ if and only if $[p_1] \preccurlyeq [p_2]$.

**Theorem 1.2.** *Definitions 1.4 and 1.7 are equivalent.*

*Proof.* (Definition 1.7 $\Rightarrow$ Definition 1.4). Let $\pi$ be a partition of the underlying set of a poset $P$ and let $\preccurlyeq$ be a partial order on $\pi$ satisfying (ii). Consider the projection map $f : P \to \pi$ which sends each element of $P$ to its block in $\pi$. Clearly, $f$ is a surjection. Since, for all $p \in P$, $f(p) = [p]$, Condition (ii) is equivalent to the fibre-coherent condition in Definition 1.2. Thus, $f$ is a fibre-coherent surjection, having $\pi$ as its set of fibres. Moreover, since $f^{-1}(B) = B$ for all $B \in \pi$, the partial order $\preccurlyeq$ satisfies (1.2). Thus, $(\pi, \preccurlyeq)$ is a regular partition of $P$ according to Definition 1.4.

(Definition 1.4 ⇒ Definition 1.7). Let $f : P \to Q$ be a fibre-coherent surjection, and let $(\pi_f, \preccurlyeq)$ be a regular partition of $P$, according to Definition 1.4. Since $f$ is surjective, $\pi_f$ is a partition of the underlying set of $P$. Consider $p_1, p_2 \in P$. By Definition 1.2, $p_1 \lesssim_{\pi_f} p_2$ if and only if $f(p_1) \leqslant f(p_2)$. By Definition 1.4, $f(p_1) \leqslant f(p_2)$ if and only if $[p_1] = f^{-1}(f(p_1)) \preccurlyeq [p_2] = f^{-1}(f(p_2))$. Thus, $p_1 \lesssim_{\pi_f} p_2$ if and only if $[p_1] \preccurlyeq [p_2]$, and (ii) is proved. Thus, $(\pi_f, \preccurlyeq)$ is a regular partition of $P$ according to Definition 1.7.

**Corollary 1.2.** *Let $(\pi, \preccurlyeq)$ be a partition of the underlying set of a poset $P$ satisfying (1.4). Then, there exists exactly one regular partition of $P$ having $\pi$ as underlying set.*

*Proof.* Let $\pi$ be a partition of the underlying set of a poset $P$ satisfying (1.4). Define the binary relation $\preccurlyeq$ on $\pi$ by prescribing that, for all $p_1, p_2 \in P$, $p_1 \lesssim_\pi p_2$ if and only if $[p_1] \preccurlyeq [p_2]$. It is straightforward to check that $\preccurlyeq$ is a partial order on $P$. By Definition 1.7, $(\pi, \preccurlyeq)$ is a regular partition of $P$. To see that it is the unique regular partition of $P$ having $\pi$ as underlying set, just observe that $\preccurlyeq$ must be a partial order satisfying Condition (ii) in Definition 1.7.

The uniqueness property of regular partitions proved in the above corollary does not hold, in general, for monotone partitions; cf. Figure 1.2, which shows three distinct monotone partitions of a given poset $P$ having the same underlying set.

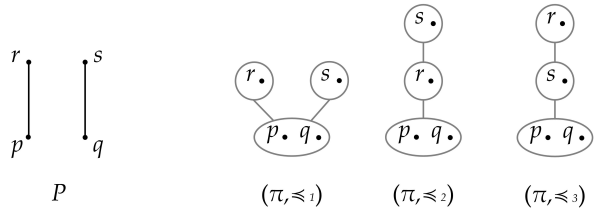

**Fig. 1.2** Distinct monotone partitions with the same support $\pi$.

Given a poset $P$ and a subset $S \subseteq P$, the *lower set* generated by $S$ is

$$\downarrow S = \{p \in P \mid p \leqslant s, \text{ for some } s \in S\}.$$

Analogously, the *upper set* generated by $S$ is $\uparrow S = \{p \in P \mid p \geqslant s, \text{ for some } s \in S\}$. When $S$ is a singleton $\{s\}$, we write $\downarrow s$ for $\downarrow \{s\}$, and $\uparrow s$ for $\uparrow \{s\}$.

**Definition 1.8 (Open partition).** An *open partition* of a poset $P$ is a poset $(\pi, \preccurlyeq)$ where

(i) $\pi$ is a partition of the underlying set of $P$ such that for each $B \in \pi$ there exist $B_1, \ldots, B_k \in \pi$ satisfying [3]

---

[3] Here and in the following lower and upper sets are always relative to the order $\leqslant$ of the poset.

$$\uparrow B = B_1 \cup \cdots \cup B_k \,. \tag{1.5}$$

(ii) for each $B, C \in \pi$, $B \preccurlyeq C$ if and only if there are $p_1 \in B$, $p_2 \in C$ such that $p_1 \leqslant p_2$.

**Theorem 1.3.** *Definitions 1.5 and 1.8 are equivalent.*

*Proof.* (Claim). If $\pi$ satisfies (i), then, for each $B, C \in \pi$, $C \subseteq \uparrow B$ if and only if there are $p_1 \in B$, $p_2 \in C$ such that $p_1 \leqslant p_2$.

Whenever $p_1 \leqslant p_2$, the block $C$ intersects the upper set of the block $B$. By (1.5), $C$ must be entirely contained in $\uparrow B$. The converse is trivial.

(Definition 1.8 $\Rightarrow$ Definition 1.5). Let $\pi$ be a partition of the underlying set of a poset $P$ satisfying (i) and let $\preccurlyeq$ be a partial order on $\pi$ satisfying (ii).

Let us consider the projection map $f : P \to \pi$ which sends each element of $P$ to its block. Let $p_1, p_2 \in P$. If $p_1 \leqslant p_2$ then, by (ii), $f(p_1) = [p_1] \preccurlyeq f(p_2) = [p_2]$ and $f$ is order-preserving. Since $\pi$ does not have empty blocks, $f$ is surjective. To show $f$ is open, we consider $p_1 \in P$, and $B \in \pi$, such that $B \preccurlyeq [p_1]$. By (ii) and (Claim), $[p_1] \subseteq \uparrow B$. Thus, there exists $p_2 \in B$ such that $p_2 \leqslant p_1$. Since $f(p_2) = [p_2] = B$, $f$ is open. Therefore, $(\pi, \preccurlyeq)$ is a regular partition of $P$ according to Definition 1.5.

(Definition 1.5 $\Rightarrow$ Definition 1.8). Let $f : P \to Q$ be a surjective open map, and let $(\pi_f, \preccurlyeq)$ be an open partition of $P$, according to Definition 1.5. Suppose, by way of contradiction, that (1.5) does not hold. Thus, there exist $p_1, p_2 \in C$ such that $p_1 \in \uparrow B$, but $p_2 \notin \uparrow B$, for some $B, C \in \pi_f$. Let $f(B) = q$. Since $f$ is order-preserving, $q \in \downarrow f(p_1)$. Since $f$ is open, $q \notin \downarrow f(p_2)$, for else we would find $p \in B$ with $p \leqslant p_2$, against the hypothesis $p_2 \notin \uparrow B$. Since $f(p_1) = f(p_2)$, $q \in \downarrow f(p_1)$ and $q \notin \downarrow f(p_2)$ is a contradiction. Thus, (i) holds.

To prove (ii) consider $p_1, p_2 \in P$, and suppose $p_1 \leqslant p_2$. Since $f$ is order-preserving, $f(p_1) \leqslant f(p_2)$. By Condition (1.3) in Definition 1.5, $[p_1] = f^{-1}(f(p_1)) \preccurlyeq [p_2] = f^{-1}(f(p_2))$, and one side of (ii) is proved.

Suppose now that for some $B, C \in \pi_f$, $B \preccurlyeq C$. Let $q_1 = f(B)$ and $q_2 = f(C)$. By Condition (1.3), $q_1 \leqslant q_2$ in $Q$. Since $f$ is surjective, fibres are not empty, and there exists $p_1 \in B$. Moreover, since $f$ is open, there exists $p_2 \in f^{-1}(q_1) = C$ such that $q_2 \leqslant q_1$. We have so proved (ii), and the proof is complete. Thus, $(\pi_f, \preccurlyeq)$ is an open partition of $P$ according to Definition 1.8.

Next we prove that each open partition is solely determined by its underlying set.

**Corollary 1.3.** *Let $P$ be a poset, and let $\pi$ be a partition of the underlying set of $P$ satisfying* (i) *in Definition 1.8. Then, there exists a unique partial order $\preccurlyeq$ on $P$ such that $(\pi, \preccurlyeq)$ is an open partition of $\pi$.*

*Proof.* (Claim). For each $B, C \in \pi$, $C \subseteq \uparrow B$ if and only if there are $p_1 \in B$, $p_2 \in C$ such that $p_1 \leqslant p_2$.

Whenever $p_1 \leqslant p_2$, the block $C$ intersects the upper set of the block $B$. By (1.5), $C$ must be entirely contained in $\uparrow B$. The converse is trivial.

Let $\preccurlyeq$ be the binary relation on $\pi$ satisfying (ii) in Definition 1.8. Clearly, $\preccurlyeq$ is uniquely determined. We need to show that $\preccurlyeq$ is a partial order on $\pi$. Reflexivity and transitivity of $\preccurlyeq$ hold trivially. To show antisymmetry, let $B, C \in \pi$ be such that $B \preccurlyeq C$ and $C \preccurlyeq B$. Let $b_1 \in B$. Since $C \preccurlyeq B$, by (Claim), there exists $c_1 \in C$ such that $c_1 \leqslant b_1$. Since $B \preccurlyeq C$ there exists $b_2 \in B$ such that $b_2 \leqslant c_1 \leqslant b_1$. We can construct in this way an infinite chain $\cdots \leqslant c_i \leqslant b_i \leqslant \cdots \leqslant b_2 \leqslant c_1 \leqslant b_1$. Since $P$ is finite, we can find an element $b_s \in B$ which occurs in the chain more then once, and, by construction, an element $c_t \in C$ such that $b_s \leqslant c_t \leqslant b_s$. Therefore, $b_s = c_t$. Since $\pi$ is a set partition, and its blocks are pairwise disjoints, we obtain $B = C$. Thus, the relation $\preccurlyeq$ is antisymmetric; therefore it is a partial order on $\pi$.

We close this section with an example.

*Example 1.1.* We refer to Figure 1.3, and consider the poset $P$.

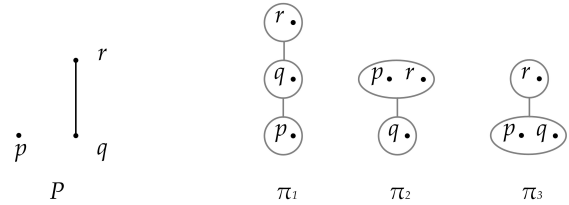

**Fig. 1.3** Example 1.1.

One can easily check, using the characterisations of poset partitions provided in Definitions 1.6, 1.7, and 1.8, that the following hold.

- $\pi_1$ is a monotone partition of $P$, but it is not regular.
- $\pi_2$ is a regular partition of $P$, but it is not open.
- $\pi_3$ is an open partition of $P$.

## 1.5 Partitions Induced by Quasiorders

A quasiorder relation $\lesssim$ on a set $A$ induces on $A$ an equivalence relation $\equiv$ defined as

$$x \equiv y \text{ if and only if } x \lesssim y \text{ and } y \lesssim x, \text{ for any } x, y \in A. \qquad (1.6)$$

The set $\pi$ of equivalence classes of $\equiv$ is a partition of $A$.

**Notation.** In the following we denote by $[x]_\lesssim$ the equivalence class (the block) of the element $x$ induced by the quasiorder $\lesssim$ via the equivalence relation defined in (1.6).

Further, the quasiorder $\lesssim$ induces on $\pi$ a partial order $\preccurlyeq$ defined by

$$x \lesssim y \text{ if and only if } [x]_\lesssim \preccurlyeq [y]_\lesssim, \text{ for any } x, y \in A. \tag{1.7}$$

We call $(\pi, \preccurlyeq)$ *the poset of equivalence classes induced by* $\lesssim$.

This correspondence allows us to give a further definition of monotone, regular, and open partition of a poset by means of quasiorders.

**Definition 1.9 (Monotone partition).** A *monotone partition* of a poset $(P, \leqslant)$ is the poset of equivalence classes induced by a quasiorder $\lesssim$ on $P$ extending $\leqslant$, *i.e.* such that $\leqslant \,\subseteq\, \lesssim$.

**Theorem 1.4.** *Definitions 1.6 and 1.9 are equivalent.*

*Proof.* (Definition 1.6 $\Rightarrow$ Definition 1.9). Let $\pi$ be a partition of the underlying set of a poset $P$, and let $\preccurlyeq$ be a partial order on $\pi$ satisfying Condition (ii) in Definition 1.6. Consider the relation $\lesssim$ on $P$ defined by

$$p_1 \lesssim p_2 \text{ if and only if } [p_1]_\pi \preccurlyeq [p_2]_\pi, \text{ for each } p_1, p_2 \in P. \tag{1.8}$$

One can easily check that $\lesssim$ is reflexive and transitive, and thus it is a quasiorder on $P$. Moreover, $[p_1]_\pi = [p_2]_\pi$ if and only if $[p_1]_\pi \preccurlyeq [p_2]_\pi$ and $[p_2]_\pi \preccurlyeq [p_1]_\pi$. Thus, by (1.8), $[p_1]_\pi = [p_2]_\pi$ if and only if $p_1 \lesssim p_2$ and $p_2 \lesssim p_1$. By (1.6), $[p_1]_\pi = [p_2]_\pi$ if and only if $[p_1]_\lesssim = [p_2]_\lesssim$. Therefore, $\pi$ coincide with the set of equivalence classes induced by $\lesssim$. Moreover, by (1.8) and (1.7), $\preccurlyeq$ coincide with the partial order on $\pi$ induced by $\lesssim$. Thus, $(\pi, \preccurlyeq)$ is the poset of equivalence classes of $P$ induced by $\lesssim$.

Suppose now that $p_1 \leqslant p_2$, for some $p_1, p_2 \in P$. By (ii) in Definition 1.6 we have $[p_1]_\pi \preccurlyeq [p_2]_\pi$. By (1.8), $p_1 \lesssim p_2$. Thus, $\lesssim$ extends $\leqslant$, and one side of the statement is proved.

(Definition 1.9 $\Rightarrow$ Definition 1.6). Let $(P, \leqslant)$ be a poset, let $\lesssim$ be a quasiorder on $P$ extending $\leqslant$, and let $(\pi, \preccurlyeq)$ be the poset of equivalence classes induced by $\lesssim$. Since $\lesssim$ extends $\leqslant$, for each $p_1, p_2 \in P$, $p_1 \leqslant p_2$ implies $p_1 \lesssim p_2$. Moreover, by (1.7), $p_1 \lesssim p_2$ implies $[p_1] \preccurlyeq [p_2]$. Thus, Condition (ii) in Definition 1.6 hold. We obtain that $(\pi, \preccurlyeq)$ is a monotone partition of $P$ according to Definition 1.6.

For what regular partitions are concerned, the definition in terms of quasiorders can be formulated as follows.

**Definition 1.10 (Regular partition).** A *regular partition* of a poset $(P, \leqslant)$ is the poset of equivalence classes induced by a quasiorder $\lesssim$ on $P$ extending $\leqslant$ and satisfying

$$\lesssim \,=\, \text{tr}(\lesssim \setminus \rho), \tag{1.9}$$

where $\text{tr}(R)$ denotes the transitive closure of the relation $R$, and $\rho$ is a binary relation defined by

$$\rho = \{(x, y) \in P \times P \mid x \lesssim y, x \not\lesssim y, y \not\lesssim x\}.$$

**Theorem 1.5.** *Definitions 1.7 and 1.10 are equivalent.*

*Proof.* (Definition 1.7 $\Rightarrow$ Definition 1.10). Let $\pi$ be a partition of the underlying set of a poset $P$, and let $\preccurlyeq$ be a partial order on $\pi$ satisfying Condition (ii) in Definition 1.7. Consider the relation $\lesssim$ on $P$ defined by

$$p_1 \lesssim p_2 \text{ if and only if } [p_1]_\pi \preccurlyeq [p_2]_\pi, \text{ for each } p_1, p_2 \in P. \qquad (1.10)$$

Since $(\pi, \preccurlyeq)$ is a monotone partition of $P$, proceeding as in the first part of the proof of Theorem 1.4, we obtain that $(\pi, \preccurlyeq)$ is the poset of equivalence classes induced by $\lesssim$, and that $\lesssim$ extends $\leqslant$. Moreover, by Condition (ii) in Definition 1.7, $\lesssim$ coincides with $\lesssim_\pi$.

It remains to prove that $\lesssim$ satisfies (1.9). Let $p_1, p_2 \in P$ be such that $p_1 \lesssim p_2$. If $(p_1, p_2) \notin \rho$, then $(p_1, p_2) \in \text{tr}(\lesssim \setminus \rho)$, trivially. Let then $(p_1, p_2) \in \rho$, that is $p_1 \lesssim p_2$, $p_2 \not\lesssim p_1$, $p_1 \not\leqslant p_2$. Since $\lesssim$ coincides with $\lesssim_\pi$, there exists a sequence $p_1 = x_0, y_0, x_1, y_1, \ldots, x_n, y_n = p_2$ satisfying Conditions (1) and (2) in Definition 1.1. By Condition (1) in Definition 1.1, and by (1.6), for each $i \in \{0, \ldots, n\}$, $x_i \lesssim y_i$ and $y_i \lesssim x_i$ hold. Thus, $(x_i, y_i) \notin \rho$. By Condition (2) in Definition 1.1, and since $\lesssim$ extends $\leqslant$, for each $i \in \{0, \ldots, n-1\}$, we have that $y_i \lesssim x_{i+1}$. Thus, $(y_i, x_{i+1}) \notin \rho$. Summing up, no pair of adjacent element in the sequence $x_0, y_0, x_1, y_1, \ldots, x_n, y_n$ belongs to $\rho$. Since each of these pairs belong to $\lesssim$, $(p_1, p_2) \in \text{tr}(\lesssim \setminus \rho)$. Thus, (1.9) holds, and the first part of the statement is settled.

(Definition 1.10 $\Rightarrow$ Definition 1.7). Let $(P, \leqslant)$ be a poset, let $\lesssim$ be a quasiorder on $P$ extending $\leqslant$ and satisfying (1.9) and let $(\pi, \preccurlyeq)$ be the poset of equivalence classes induced by $\lesssim$.

(Claim 1). $p_1 \lesssim_\pi p_2$ implies $p_1 \lesssim p_2$, for each $p_1, p_2 \in P$.

Suppose that $p_1 \lesssim_\pi p_2$, for some $p_1, p_2 \in P$. Thus, there exists a sequence $p_1 = x_0, y_0, x_1, y_1, \ldots, x_n, y_n = p_2 \in P$ satisfying Conditions (1) and (2) in Definition 1.1. By Condition (1), for each $i \in \{0, \ldots, n\}$, $[x_i]_\pi = [y_i]_\pi$. Thus, by (1.7), $x_i \lesssim y_i$. By Condition (2), for each $i \in \{0, \ldots, n-1\}$, $y_i \leqslant x_{i+1}$. Since, by hypothesis, $\lesssim$ extends $\leqslant$, we have $y_i \lesssim x_{i+1}$. Therefore, whenever $p_1 \lesssim_\pi p_2$, we have $p_1 = x_0 \lesssim y_0 \lesssim \cdots \lesssim x_n \lesssim y_n = p_2$. By transitivity $p_1 \lesssim p_2$.

(Claim 2). $p_1 \lesssim p_2$ implies $p_1 \lesssim_\pi p_2$, for each $p_1, p_2 \in P$.

Let $p_1 \lesssim p_2$, for some $p_1, p_2 \in P$. We shall analyze three different cases, covering all the possibilities for $p_1$ and $p_2$.

Case (a). $p_2 \lesssim p_1$. Since $p_1 \lesssim p_2$ and $p_2 \lesssim p_1$, by (1.6), we have $[p_1]_\pi = [p_2]_\pi$. Thus, the sequence $p_1, p_2$ satisfies Conditions (1) and (2) in Definition 1.1, with respect to $\pi$. We obtain $p_1 \lesssim_\pi p_2$.

Case (b). $p_2 \not\lesssim p_1$, and $p_1 \leqslant p_2$. Since the sequence $p_1, p_1, p_2, p_2$ satisfies Conditions (1) and (2) in Definition 1.1, we have $p_1 \lesssim_\pi p_2$.

Case (c). $p_2 \not\lesssim p_1$, $p_1 \not\leqslant p_2$, that is, $(p_1, p_2) \in \rho$. Let $R = \lesssim \setminus \rho$. Since $p_1 \lesssim p_2$, the pair $(p_1, p_2)$ arises in $\lesssim$ from the transitive closure of the binary relation $R$. Thus, there exists a sequence $p_1 = z_0 \lesssim z_1 \lesssim \cdots \lesssim z_r = p_2$ of elements of $P$ such that, for all $i \in \{0, \ldots, r-1\}$, $z_i \, R \, z_{i+1}$. For each of this pair of elements, either Case (a) or Case (b) apply. Thus, $z_i \lesssim_\pi z_{i+1}$. Since $\lesssim_\pi$ is transitive, we obtain immediately $p_1 \lesssim_\pi p_2$.

In any case, whenever $p_1 \lesssim p_2$, we have $p \lesssim_\pi q$, and the claim is settled.

By (Claim 1) and (Claim 2), the quasiorders $\lesssim$ and $\lesssim_\pi$ coincide, and Condition (ii) in Definition 1.7 is trivially verified. We obtain that $(\pi, \leqslant)$ is a regular partition of $P$ according to Definition 1.7.

*Remark 1.3.* Let $(\pi, \leqslant)$ be the monotone partition induced by a quasiorder $\lesssim$. By the construction given in the proof of Theorem 1.5, we infer that $(\pi, \leqslant)$ is a regular partition if and only if $\lesssim$ coincides with $\lesssim_\pi$.

In the case of open partitions, the definition in terms of quasiorders is as follows.

**Definition 1.11 (Open partition).** An *open partition* of a poset $(P, \leqslant)$ is the poset of equivalence classes induced by a quasiorder $\lesssim$ on $P$ extending $\leqslant$ and such that, for every $p, q \in P$,

$$\text{if } p \lesssim q, \text{ then there exists } p' \in P \text{ such that } p \lesssim p', \ p' \lesssim p, \text{ and } p' \leqslant q. \quad (1.11)$$

**Theorem 1.6.** *Definitions 1.8 and 1.11 are equivalent.*

*Proof.* (Definition 1.8 $\Rightarrow$ Definition 1.11). Let $(\pi, \leqslant)$ be an open partition of a poset $P$, satisfying thus Conditions (i) and (ii) in Definition 1.8. Consider the relation $\lesssim$ on $P$ defined by

$$p_1 \lesssim p_2 \text{ if and only if } [p_1]_\pi \leqslant [p_2]_\pi, \text{ for each } p_1, p_2 \in P. \quad (1.12)$$

Since $(\pi, \leqslant)$ is a monotone partition of $P$, proceeding as in the first part of the proof of Theorem 1.4, we obtain that $(\pi, \leqslant)$ is the poset of equivalence classes induced by $\lesssim$, and that $\lesssim$ extends $\leqslant$. Let $p \lesssim q$ for some $p, q \in P$. By (1.12), $[p]_\pi \leqslant [q]_\pi$. By Conditions (i) and (ii) in Definition 1.8, $[p]_\pi \leqslant [q]_\pi$ implies $[q]_\pi \subseteq \uparrow [p]_\pi$. Thus, there exists $p' \in [p]_\pi$ such that $p' \leqslant q$, and (1.11) is satisfied. Therefore, $(\pi, \leqslant)$ is an open partition of $P$ according to Definition 1.11.

(Definition 1.11 $\Rightarrow$ Definition 1.8). Let $(\pi, \leqslant)$ be the poset of equivalence classes induced by $\lesssim$.

(Claim). Let $B, C \in \pi$. Then, $B \leqslant C$ if and only if $C \subseteq \uparrow B$.

($\Leftarrow$) Suppose $C \subseteq \uparrow B$. Since blocks are nonempty, there exist $p \in B$ and $q \in C$ such that $p \leqslant q$. Since $\lesssim$ extends $\leqslant$, we have $p \lesssim q$. Hence, by (1.7), $B \leqslant C$.

($\Rightarrow$) Suppose $B \leqslant C$. By (1.7), for each $q \in C$, $p \in B$, we have $p \lesssim q$. Moreover, by (1.11), there exists $p' \in B$ such that $p' \leqslant q$. Thus, for each $q \in C$ there exists $p' \in B$ such that $p' \leqslant q$. In other words, $C \subseteq \uparrow B$.

Conditions (ii) in Definition 1.8 follows immediately from (Claim). Let $B, C \in \pi$, and suppose that $p \leqslant q$, for some $p \in B$, $q \in C$. Since $\lesssim$ extends $\leqslant$, we have $p \lesssim q$. Moreover, by (1.7), $B \leqslant C$. By (Claim), $C$ is entirely contained in $\uparrow B$. We have so proved that whenever a block $C$ intersects $\uparrow B$, the block $C$ is entirely contained in $\uparrow B$. Thus, $\uparrow B$ can be written as a union of blocks of $\pi$, as in Conditions (i) in Definition 1.8, and the theorem is proved.

## 1.6 Further Work

The further development of a theory of partitions of finite posets can follow several research direction. One of these concerns enumeration problems. It is easy to realize that enumerating poset partitions is never a simple problem, except in trivial cases – *e.g.* counting monotone, regular, and open partitions of chains, or of trivially ordered posets. Indeed, as we have shown, the enumeration of poset partitions is tightly related to the problem of counting the number of quasiorders on a finite set of points. Special cases may be tractable; we plan to address the issue in future work, following the relatively simple cases investigated in [3].

Another direction the we plan to follow in our future research involves the investigation of the ordered structure of all poset partitions of a poset. In [3], we proved that the classes of all monotone and regular partitions of a poset form a lattice, and we analysed its first properties.

Applications of the notions of monotone, regular, and open partitions can also be found in the study of distributive lattices, and Heyting algebras. Indeed, we recall here the well-known fact that the category Pos is dually equivalent to the category of finite bounded distributive lattices with their $\{0,1\}$-preserving lattice homomorphisms (for details see, *e.g.*, [5]). In the dual category of finite distributive lattices our notions of monotone and regular partition correspond precisely to the notions of sublattice and regular sublattice (see [3, Chapter 5]), respectively, of a distributive lattice.

Concerning open partitions, the category OPos can be proved to be dually equivalent to the category of finite Heyting algebras with their homomorphisms (for details on Heyting algebras, see, *e.g.*, [6]). In the dual category of finite Heyting algebras our notion of open partition corresponds precisely to the notion of subalgebra.

The above duality allows, for instance, the application of our results on open partitions to the study of the notion of probability distribution in Gödel logic (an extension of the intuitionistic propositional calculus); please see [4].